\documentclass{article}
\begin{document}
\begin{center}
{\Large{ Holographic Ricci dark energy as running vacuum}} \\[0.2in]
Paxy George and Titus K Mathew \\
Department of Physics, \\
Cochin University of Science and Technology, Kochi-22, India.
\end{center}
\begin{abstract}

Holographic Ricci dark energy has been proposed ago has faced with
problems of future singularity. In the present work we consider the
Ricci dark energy with an additive constant in it's density as
running vacuum energy. We have analytically solved the Friedmann
equations and also the role played by the general conservation law
followed by the cosmic components together. We have shown that the
running vacuum energy status of the Ricci dark energy helps to
remove the possible future singularity in the model. The additive
constant in the density of the running vacuum played an important
role, such that, without that, the model predicts either eternal
deceleration or eternal acceleration. But along with the additive
constant, equivalent to a cosmological constant, the model predicts
a late time acceleration in the expansion of the universe, and in
the far future of the evolution it tends to de Sitter universe.
\end{abstract}

\section{Introduction}
\label{sec:intro}

      The accelerated expansion of the current universe is a confirmed phenomenon as per the
present observations.
      The observational data from the supernova type Ia (SNeIa) \cite{3,4}, cosmic microwave background
\cite{5}, large scale structure (LSS) \cite{6}, baryon acoustic
oscillations \cite{7} and weak lensing \cite{8} are the strong
supports for this. The reason for this accelerated expansion is
argued to be due to the presence of gravitationally repulsive energy
component known as "dark energy" (DE).The nature and evolution of
dark energy is still not clear inspite of the numerous speculation
existing in the current literature.

The simplest model for dark energy is by taking it as the vacuum
energy or cosmological constant, lead to the standard $\Lambda$CDM
model of the universe. This approach faces coincidence problem and
fine tunning problem \cite{9}. The coincidence problems means that,
even though the cosmic evolution of both dark matter and dark energy
are  different their densities of the same order in the present
universe, a thing which is not accounted by the standard models of
the universe. The fine tuning problems is that, the theoretically
predicted value of the cosmological constant is several orders
higher than the observed value. These problems motivated the
consideration of various dynamical dark energy models like
quintessence \cite{10,11}, kessence \cite{12}, and the Chaplygin gas
model \cite{13}, in which the dark energy density is evolving with
time. Another approach in explaining the recent acceleration of the
universe, is by modifying the left hand side of the Einstein
equation, i.e., the geometry of the space time, which are generally
called as modified gravity theories. The dark energy models based on
modified gravity are the so-called f(R) gravity \cite{14}, f(T )
gravity \cite{15}, Gauss– Bonnet theory \cite{16}, Lovelock gravity
\cite{17},scalar–tensor theories \cite{18} etc.

   A class of models which were proposed to alleviate the problems of the $\Lambda$CDM model are the
dynamical vacuum energy models. Recently much attention has been paid in these models in which dark energy treated as a  time varying vacuum in which
the density is varying as the universe expands, but
   equation of state stays constant around -1, and is often termed as running vacuum energy\cite{19,20,21}. The
   main motivation for this approach is arising from vacuum energy
   predicted by quantum field theory in curved space-time derives from the
   renormalization group. The vacuum energy predicted from such
   theories, posses constant equation of state but their density is
   varying during the evolution.
   The evolution of such a running vacuum energy is considered in reference \cite{22}, where the authors have
   shown that, the recent acceleration of the universe can be
   explained with varying density contain an extra constant term, and corresponding equation of state is stands constant
   around -1. The model is free from future singularity under certain conditions on the model parameters. The recent
   Plank observations, however point towards the possibility of a
   phantom equation of state for the dark energy that, $\omega =
   -1.49^{+0.65}_{-0.57}$\cite{34}, but the wide error bars
   makes the phantom nature still doubtful. On the other hand
   the WMAP observations point towards a value around, $\omega \sim
   $-0.93\cite{35}.  But the latest result combining WMAP, BAO, CMB,$ H_{0}$ is found to
   be, $\omega = -1.073^{+0.090}_{-0.089}$\cite{36}.
    Both these results a value around -1 for the equation of state,which can be taken as a concordance one.
   Moreover the success of $\Lambda$CDM model
   which treat dark energy as the vacuum constant also pointing
   towards a constant equation of state around $-1.$
   In addition to the constant equation of state, the evolutionary nature of the density in running vacuum energy models may
   safeguard the coevolution of the dark sectors.

   In the present work we considered Ricci dark energy as the
   running vacuum energy. This form of dark energy is derived from
   holographic principle\cite{28,29}. The terms in Ricci dark energy is almost
   similar to that  considered in paper\cite{23}. We append the
   usual form of dark energy with an additional constant term in the
   density. The paper is organized as follows. After a brief introduction in section \ref{sec:intro}, we discuss running vacuum energy, and Ricci dark energy as running vacuum
energy in section \ref{sec:sec2}. This section also consists of our analysis of the evolution of Hubble parameter, density function and also other relevant cosmological
parameter. We conclude in section \ref{sec:sec3}.

\section{Holographic Ricci dark energy as running vacuum energy}
\label{sec:sec2}
According to holographic principle\cite{25} black holes are the
maximally entropic objects of a given region.  Hence the total
degrees of freedom is bounded by the surface area in Planck units.
 Cohen et al.\cite{26} have suggested that the total vacuum
energy in a region of size L should not exceed the mass of a black
hole of the same size, $L^{3}\rho_{vac} \leq L M_{p}^{2}$ where
$\rho_{vac}$ is the quantum zero point energy density and
$M_{p}^{2}=1/ \sqrt{8\pi G}$ is the reduced Plank mass. Based on
this result, Li\cite{27} proposed the holographic dark energy as
 $\rho_{\Lambda}=3c^{2}M_{p}^{2}L^{-2}$ where $c^{2}$ is a dimensionless constant, and $L$ is often called as IR cutoff.
Hsu\cite{28} showed that the holographic dark energy with Hubble
horizon as IR cutoff does not leads to an accelerating universe.
Hence in order to explain an accelerating universe Li suggested the
future event horizon as IR cutoff instead of Hubble horizon and
particle horizon. Later studies found that in taking  future horizon
as IR cutoff will lead to causality violation. Hence Gao et
al\cite{29} proposed the holographic Ricci dark energy model in
which holographic dark energy is proportional to the Ricci scalar.
\begin{equation}
\rho_{\Lambda}=-\frac{\alpha}{16\pi}R
\end{equation}
where R is the Ricci scalar curvature $
R=-6(\dot{H}+2H^{2}+\frac{k}{a^{2}})$ \cite{30} and $\alpha$ is a
constant to be determined, $H=\dot{a}/a$, the Hubble parameter,k is
the curvature parameter and a is the scale factor. Later modified
form of holographic Ricci dark energy were also been
studied\cite{31,32,33}.The general behavior of equation of state in
these models is that it can assume values greater than or less than
$-1$ corresponds to quintessence or phantom nature depending on the
model parameter.

Running vacuum energy has got it's original form as\cite{19},
\begin{equation}\label{eqn:rv1}
 \rho_{\Lambda}=n_0+n_1H^2+\mathcal{O}(H^4).
\end{equation}
This is the vacuum energy in quantum field theory in curved
space-time originating from the renormalization group equation in
reference \cite{19}. Only even powers of $H$ are allowed to maintain
the general covariance. Since higher order powers are much smaller,
they are often neglected. The constant $n_0$ is often replaces the
role of the cosmological constant and $n_1$ is a dimensionless
parameter given by\cite{19},
\begin{equation}
 n_1=\frac{3\nu}{8\pi}M_P^2
\end{equation}
where $\nu$ is,
\begin{equation}
 \nu=\frac{1}{6\pi}\sum_i B_i \frac{M_i^2}{M_P^2}
\end{equation}
with $B_i$ as the coefficients computed from the quantum loop contributions of the fields with masses $M_i.$ The
value of $\nu$ is such that $|{\nu}|<<1$ hence the running vacuum is very close to the true cosmological constant. However small may be the value
of $\nu$ it gives a running status for the dark energy.

 In this work we
   considered  holographic Ricci dark energy as running vacuum energy which takes the form
 \begin{equation}\label{eqn:DE}
   \rho_{\Lambda}(H,\dot{H})= 3\beta M_{p}^{2}(\dot{H}+2H^{2})+ M_{p}^{2}\Lambda_{0}
   \end{equation}
    In
   this equation only the first part on the right hand side
   corresponds to the conventional Ricci dark energy. The addition
   of the second term proportional to $\Lambda_0$ which is a constant, is aiming at to guarantee a
   transition from an early deceleration to an acceleration phase
   in the expansion history of the universe. In dealing with
   entropic dark energy as dynamical vacuum energy in reference
   \cite{23}, the authors considered a similar addition of a constant
   term to energy density, while in the absence of such a term the universe may undergo either eternal deceleration or acceleration.
In the above equation for $\beta$=0, the dark energy density reduces
to the cosmological constant as in the case of the original running
vacuum equation (\ref{eqn:rv1}). Unlike the original running vacuum
form, there appears a term $\dot{H}$ in the present density
equation. But this term will not make generally different from the
original form, because, first of all, both $H^2$ and $\dot H$ are of
same dimension even though they represents different degrees of
freedom. Secondly these terms are related through $\dot H =
-(1+q)H^2,$ where $q$ is the deceleration parameter. During
different stages of the cosmic evolution, the $q$ parameter appears
roughly a constant. For instance, during radiation dominated phase,
the parameter $q=1,$ while for matter dominated phase, $q=0.5.$
Hence $\dot H \sim H^2$ in these cosmic phases. So during these
phases there exist a correspondence between the parameter $n_1$ or
$\nu$ in the original running vacuum and the $\beta$ parameter in
the holographic Ricci running vacuum. However in the later phase of
cosmic acceleration the $q$ parameter is undergoing a variation, but
due to the smallness of the $\beta$ parameter the $\dot H$ term
behave almost like the $H^2$ term.

   The Friedman equations with radiation, non-relativistic matter and dark energy as cosmic components is given by Eqn
   (\ref{eqn:DE}) are,
\begin{equation} \label{eqn:F1}
   \frac{\dot{a}^2}{a^{2}}=\frac{1}{3M_{p}^{2}}\left(\rho_r+\rho_{m}+\rho_{\Lambda}(H,\dot{H})\right)
    \end{equation}
where $\rho_r$ is the radiation density and $\rho_m$ is density of
the non-relativistic matter. The equation representing the cosmic
acceleration is given as,
 \begin{equation}\label{eqn:F2}
   \frac{\ddot{a}}{a}=-\frac{1}{6M_{p}^{2}}((\rho_i+3p_i)+2p_{\Lambda}(H,\dot{H})).
   \end{equation}
where $\rho_i$ and $p_i$ together representing the density and pressure of the radiation and matter components.
Since the dark energy is of the nature of dynamical vacuum energy,
it would satisfy the condition,
\begin{equation}\label{eqn:ded}
   \rho_{\Lambda}(H,\dot{H})= -p_{\Lambda}(H\dot{H})= 3\beta M_{p}^{2}(\dot{H}+2H^{2})+ M_{p}^{2}\Lambda_{0}
   \end{equation}

   An important feature of running vacuum energy is it's decay. For the covariance nature of the theory, Bianchi identity must be satisified\cite{38},
which insure the covariance nature of the equations,
 and further implies that, it is the total energy density
of the entire system of universe is conserved \cite{24}, that is a
separate conservation law for each individual component is not
followed. This effectively take
    account of the transfer of energy between running vacuum
   energy density and other components presents in the universe.
   The conservation equation then takes the form
   \begin{equation}
   \dot{\rho}_{m}+\dot{\rho}_{r}+\dot{\rho}_{\Lambda}+3H(\rho_{m}+\frac{4}{3}\rho_{r})=0.
   \end{equation}
After substituting the time derivative of the dark energy density, the above equation become,
\begin{equation}\label{eqn:con123}
 \dot{\rho}_m+\dot{\rho}_r -\frac{3}{2}\beta \left(\dot{\rho}_m + \frac{4}{3}\dot{\rho}_r\right)=-3H\left(1-2\beta\right) \left(\rho_m+\frac{4}{3}\rho_r\right)
\end{equation}
This means that even in the case allowing an arbitrary interaction
$Q(t)$ between matter and radiation, then any solution of the
equations of the form

 \begin{equation}
   \dot{\rho_{m}}=-3H\frac{(1-2\beta)}{(1-\frac{3}{2}\beta)}\rho_{m}+Q(t)
   \end{equation}
  \begin{equation}
 \dot{\rho_{r}}=-4H\rho_{r}-Q(t).
 \end{equation}
are simultaneously be the solutions of the equation (\ref{eqn:con123}). In stages where either matter or radiation dominated
the densities, the equation (\ref{eqn:con123}) will be satisfied, that is $Q(t)\to 0.$
Hence a simplest version of this is to assume that the total
conservation equation reduces to a set of decoupled equations with
$Q=0$ at all times, which may be the only way to introduce arbitrary
number of cosmic components. Then the evolution of matter during
matter dominated era and radiation during radiation dominated era
are become
 \begin{equation}\label{eqn:mat}
 \rho_{m}=\rho_{m_{0}}a^{-3\xi_{m}}
 \end{equation}
 \begin{equation}\label{eqn:rad}
  \rho_{r}=\rho_{r_{0}}a^{-4}
\end{equation}
where $\xi_{m}=\frac{(1-2\beta)}{(1-\frac{3}{2}\beta)}$,
$\rho_{m_{0}}$ and $\rho_{r_{0}}$ are energy density of matter and
radiation at the present time respectively. Here it should be noted that, due to the decay of running vacuum, the evolution of the
matter density is modified but the radiation follows the conventional behavior that, $\rho_r \propto a^{-4}.$ This may indicate
 that the running vacuum is substantially coupled with the matter than radiation in the present model. By substituting
Eqns(\ref{eqn:mat}), (\ref{eqn:rad}) and (\ref{eqn:ded}) in
Eqn(\ref{eqn:F1}) we can finally reached to

\begin{equation}\label{eqn:doth}
\frac{\dot{h}}{H_{0}}=\frac{-1}{\beta}[\Omega_{m_{0}}e^{-3\xi_{m}x}+\Omega_{r_{0}}e^{-4x}+(2\beta-1)
h^{2}+\frac{\Lambda_0}{3H_{0}^{2}}]
\end{equation}
where $\Omega_{m_{0}}=\frac{\rho_{m_{0}}}{3 M_{p}^{2}H_{0}^{2}}$,
 $\Omega_{r_{0}}=\frac{\rho_{r_{0}}}{3 M_{p}^{2}H_{0}^{2}}.$

By changing the variable from time $t$ to $x=lna$ the differential
equation can be expressed as
\begin{equation}
\frac{dh^{2}}{dx}=\frac{-2}{\beta}[\Omega_{m_{0}}e^{-3\xi_{m}x}+\Omega_{r_{0}}e^{-4x}+(2\beta-1)
h^{2}+\frac{\Lambda_0}{3H_{0}^{2}}]
\end{equation}
The solution of the above equation is
\begin{equation}\label{eqn:h}
h^{2}=\frac{\Omega_{m_{0}}}{\xi_{m}}e^{-3\xi_{m}x}+\Omega_{r_{0}}e^{-4x}+\frac{\Lambda_0}{3(1-2\beta)H_{0}^{2}}+\Omega_{\beta}e^{-(4-\frac{2}{\beta})x}
\end{equation}
where $\Omega_{\beta}$ is the integration constant. In Eqn
(\ref{eqn:h}) the first term is proportional to matter density,
second term represent the energy density of radiation and the last
two terms together represents running vacuum energy density. This
shows that in the far future running vacuum energy density will be
the dominating component of the universe\cite{23}, especially at
$z\to -1$ (see that $e^{-x} = (1+z)$)the hubble parameter become a
constant. ie $h^2 \to \frac{\Lambda_0}{3(1-2\beta)H_{0}^{2}}$. It is
trivial that for $h^2$ is to positive definite in the future the
parameter $\beta<1/2.$
 The first time derivative for the Hubble
parameter is obtained
 by substituting
Eq.(\ref{eqn:h}) in eqn.(\ref{eqn:doth})
\begin{equation}
\frac{\dot{h}}{H_{0}}=\frac{-3}{2}\Omega_{m_{0}}e^{-3\xi_{m}x}-2\Omega_{r_{0}}e^{-4x}-\frac{(2\beta-1)}{\beta}\Omega_{\beta}e^{-(4-\frac{2}{\beta})x},
\end{equation}
which implies that as $z \to -1$ the time derivative, $\dot h \to 0,$ otherwise implies $h \to constant.$
The running vacuum energy density is then obtained by substituting
Eq(\ref{eqn:doth}) and Eq(\ref{eqn:h}) in Eq(\ref{eqn:DE}) as
\begin{equation}\label{eqn:tot de dens}
\rho_{\Lambda}=\frac{3\beta
M_{p}^{2}H_{0}^{2}}{2-4\beta}\Omega_{m_{0}}e^{-3\xi_{m}x}+
3M_{p}^{2}H_{0}^{2}\Omega_{\beta}e^{-(4-\frac{2}{\beta})x}
+\frac{M_{p}^{2}\Lambda_0}{1-2\beta}
\end{equation}
The obtained energy density is depending on energy density of matter
but not on the radiation density and the last two terms plays significant role in the future
evolution of the universe.It will continue as an exponentially decreasing term if $\beta>1/2.$
 However for $\beta>1/2$ the last term become negative and as $z \to -1$ the entire density
$\rho_{\Lambda}$ tends to a negative cosmological constant.
 For $\beta<1/2$ the last term will become positive definite, but the second term will eventually lead to a big rip
singularity as $z \to -1.$ The above equation corresponds to an effective cosmological constant,
\begin{equation}
 \Lambda=\frac{3\beta
H_{0}^{2}}{2-4\beta}\Omega_{m_{0}}e^{-3\xi_{m}x}+
3H_{0}^{2}\Omega_{\beta}e^{-(4-\frac{2}{\beta})x}
+\frac{\Lambda_0}{1-2\beta}.
\end{equation}
The evolution of this cosmological term depends crucially on the
first two terms. The total energy density filling the universe would
be, $\rho_{tot}=\rho_{m}+\rho_{r}+\rho_{\Lambda}$ which results in
\begin{equation}
\rho_{tot}=3M_{p}^{2}H_{0}^{2}\frac{\Omega_{m_{0}}}{\xi_{m}}e^{-3\xi_{m}x}+
3M_{p}^{2}H_{0}^{2}\Omega_{r_{0}}e^{-4x}+\frac{M_{p}^{2}\Lambda}{1-2\beta}+3M_{p}^{2}H_{0}^{2}\Omega_{\beta}e^{-(4-\frac{2}{\beta})x}
\end{equation}
This total energy density satisfies the conservation law
\begin{equation}\label{eqn:con}
\dot{\rho_{tot}}+3H(\rho_{tot}+p_{tot})=0
\end{equation}
where the total pressure, $p_{tot}=p_{m}+p_{r}+p_{\Lambda}$. Here
$p_{m}=0, p_{r}=\frac{1}{3}\rho_{r},p_{\Lambda}=-\rho_{\Lambda}$.
Then the total conservation law takes the form
\begin{equation}\label{eqn:omega}
\Omega_{\beta}e^{-(4-\frac{2}{\beta})x}=0
\end{equation}
 The above equation is satisfied if the integration constant identically vanishes, i.e. $\Omega_{\beta}=0.$ This will reduces
the running vacuum energy equation (\ref{eqn:tot de dens}) to the simple form,
\begin{equation}\label{eqn:rholambdanew}
\rho_{\Lambda}=\frac{3\beta
M_{p}^{2}H_{0}^{2}}{2(1-2\beta)}\Omega_{m_{0}}e^{-3\xi_{m}x}
+\frac{M_{p}^{2}\Lambda_0}{1-2\beta}
\end{equation}
For positive definite value for the running vacuum density, the parameter $\beta<1/2.$
For $\beta=0$ there arise possibility where the density can have negative values in the future and also effectively kills the original
expression of the dark energy density. Hence the
$\beta$ parameter can be fixed in the range $0<\beta<1/2.$ The total energy density become,
\begin{equation}
\rho_{tot}=3M_{p}^{2}H_{0}^{2}\frac{\Omega_{m_{0}}}{\xi_{m}}e^{-3\xi_{m}x}+
3M_{p}^{2}H_{0}^{2}\Omega_{r_{0}}e^{-4x}+\frac{M_{p}^{2}\Lambda_0}{1-2\beta}
\end{equation}

In the far future of the universe as $z\to -1,$ the total energy
density will be dominated by the constant term, ie
$\rho_{tot}\to\frac{M_{p}^{2}\Lambda}{1-2\beta}.$ n law itself
demands the

It is the constant addition in the dark energy density
(ie,$M_{p}^{2}\Lambda$) guarantees the transition of the universe to
the late acceleration phase, the integration constant is no longer
be essential.

The deceleration parameter, $q=-1-\frac{\dot{H}}{H^{2}}$, is
evaluated as
\begin{equation}
q=-1+\frac{3\Omega_{m_{0}}e^{-3\xi_{m}x}}{2[\frac{\Omega_{m_0}}{\xi_{m}}e^{-3\xi_{m}x}+\frac{\Omega_{\Lambda_0}}{1-2\beta}]}
\end{equation}
For a matter dominated phase deceleration parameter $q$ can be
approximated as
\begin{equation}\label{eqn:q}
q\sim-1+\frac{3}{2}\xi_m.
\end{equation}
For $\xi_m=1,$ which otherwise implies $\beta=0,$ means contribution
of Ricci dark energy is only a constant term, the deceleration\, $q \sim 0.5$ which
corresponds to the behavior of dust like matter dominated universe
which is eternally decelerating. For the case, $\xi_m$ is different
from $1$, at which dark energy consists of both the additive constant and the varying part, see equation (\ref{eqn:ded}),
the universe may be either eternally decelerating or
accelerating. For instance, when $\xi_m >0.7$ the universe will still
be matter dominated and would be eternally decelerating. On the
other hand for $\xi_m <0.7$  the universe will be dominated by dark
energy and would be eternally accelerating. These facts indicating
that without the constant term $\Omega_{\Lambda_{0}}$ in the Ricci
dark energy density  a transition from a decelerating phase to an
accelerating phase is impossible.
On the other hand if the term
$\Omega_{\Lambda_0}$ dominates then q takes the form
\begin{equation}\label{eqn:dec123}
q=-1+\frac{3}{2}\frac{\Omega_{m_0}}{\Omega_{\Lambda_{0}}}(1-2\beta)e^{-3\xi_mx}
\end{equation}
For
the case,$\xi_m=1,$ which implies $\beta=0,$ the deceleration
parameter reduces to
\begin{equation}
q=-1+\frac{3}{2}\frac{\Omega_{m_0}}{\Omega_{\Lambda_{0}}}e^{-3x}
\end{equation}
This shows that there can occur a transition from deceleration to
acceleration for any positive values of $\xi_m$ as far the dark energy is dominated by the constant term.
The equation (\ref{eqn:dec123}) also shows that in the far future of the evolution of the
universe, $z\to-1,$ at which the density is dominated by the
constant term $\Omega_{\Lambda_0}$ and also $e^{-3\xi_mx}=(1+z)^{3\xi_m x} \to 0$ , the deceleration parameter tends to the value -1.
That is the universe approach the de Sitter phase as $z\to-1$.

\section{Conclusions}
\label{sec:sec3} In this letter we consider  holographic Ricci dark
energy plus an additional constant as running vacuum energy. Due to
the decay of vacuum in to other possible , satisfying the equation
of state $\omega=-1.$  Because of the decay of running vacuum, we
have consider a general conservation law, satisfied by the entire
cosmic components together, which reveals that, the behavior of
matter density is modified as, $\rho_m=\rho_{m0} a^{-3\xi_m}$ but
the radiation follows the conventional behavior,
$\rho_r=\rho_{r0}a^{-4}.$ The Hubble parameter were evaluated and
also the total energy density. The condition of the total
conservation law on the total density, constrain the integration
constant appeared while obtaining the hubble parameter to be equal
to zero, otherwise the model will eventually leads to a future
singularity, where both density and hubble parameter are eventually
become infinity, otherwise known as big-rip. The evolution equation
of the dark energy density thus appearing will always be positive
definite only if the model parameter is in the range, $0<\beta<1/2.$
On evaluating the deceleration parameter, it is seen that, the model
leads to late acceleration, at which the running vacuum dominates
over the other cosmic components, only in the presence of the
additive constant in the equation of the running vacuum density, but
without that the model will leads to either eternal deceleration or
acceleration. In the above specified range of the model parameter
$\beta$ the model will asymptotically become de Sitter type. As a
comparison to the present analysis, it should be noted that, along
with considering the entropic dark energy as running vacuum, the
authors \cite{23}, have analyzed the general evolution of the Ricci
dark energy, in which the model will reduce to the de Sitter phase,
only if the additive constant to the density is negative. When one
consider entropic dark energy as running vacuum\cite{23}, the
evolution of the running is depends on the radiation component too
and the behavior of the radiation component itself is modified due
to the running status of the entropic dark energy. While in the
present model, first of all, the running vacuum energy is not
depending the radiation component, and secondly, the behavior of the
radiation component is not modified but follows the conventional
behavior.

\end{document}